\documentclass{ws-procs9x6}
\def\la{\langle}\def\ra{\rangle}
\def\be{\begin{eqnarray}}\def\bea{\begin{eqnarray}}
\def\ba{\begin{eqnarray}}
\def\ee{\end{eqnarray}}\def\eea{\end{eqnarray}}
\def\ea{\end{eqnarray}}
\def\ben{\begin{eqnarray}}\def\bitem{\begin{itemize}}
\def\een{\end{eqnarray}}\def\eitem{\end{itemize}}
\def\del{\partial}
\def\G0p{$G_0^\prime$}
\def\roughly#1{\mathrel{\raise.3ex\hbox{$#1$\kern-.75em%
\lower1ex\hbox{$\sim$}}}}\def\lsim{\roughly<}

\def\Tr{\rm Tr}
\def\prl{Phys. Rev. Lett.}\def\pr{Phys. Rev.}

\def\bi{\bibitem}

\begin{document}

\title{On Effective Degrees of Freedom at Chiral Restoration
and the Vector Manifestation\footnote {\uppercase{L}ecture given
at 2002 \uppercase{I}nternational \uppercase{W}orkshop on
``\uppercase{S}trong \uppercase{C}oupling \uppercase{G}auge
\uppercase{T}heories and \uppercase{E}ffective \uppercase{F}ield
\uppercase{T}heories (\uppercase{SCGT} 02)," 10 - 13
\uppercase{D}ecember 2002, \uppercase{N}agoya
\uppercase{U}niversity, \uppercase{N}agoya, \uppercase{J}apan, to
appear in {\uppercase{W}orld \uppercase{S}cientific}.}}

\author{Mannque Rho}

\address{Service de Physique Th\'eorique, CEA/DSM/SPhT,\\
Unit\'e de recherche associ\'ee au CNRS, CEA/Saclay,  \\ 91191
Gif-sur-Yvette c\'edex, France \\
E-mail: rho@spht.saclay.cea.fr\\
\& \\
School of Physics, \\ Korea Institute for Advanced Study,
\\ Seoul
130-722, Korea\\
E-mail: rho@kias.re.kr}


\maketitle

\abstracts{Recent research activities on the chiral structure of
hadronic matter near the phase transition predicted by QCD and
extensively looked for in terrestrial laboratories as well as in
satellite observatories raise the issue of whether we have fully
identified the relevant degrees of freedom involved in the
transition. In this talk, I would like to discuss a recent novel
approach to the issue based on the ``vector manifestation"
scenario discovered by Harada and Yamawaki in hidden local
symmetry theory~\cite{HY:VM,HY:PR}. For simplicity, I will
restrict myself to two extreme scenarios:  one that we shall refer
to as ``standard" in which pions are considered to be the $only$
low-lying degrees of freedom and the other that could be referred
to as ``non-standard" in which in addition to pions, other degrees
of freedom figure in the process. In particular, I shall consider
the scenario that arises at one-loop order in chiral perturbation
theory with hidden local symmetry Lagrangian consisting of pions
as well as nearly massless vector mesons that figure importantly
at the ``vector manifestation (VM)" fixed point. It will be shown
that if the VM is realized in nature, the chiral phase structure
of hadronic matter can be much richer than that in the standard
one and the chiral phase transition will be a smooth crossover:
Sharp vector and scalar excitations are expected in the vicinity
of the critical point. Some indirect indications that lend support
to the VM scenario are discussed.}

\section{Introduction}
In describing the chiral restoration transition at the critical
temperature $T_c$ and/or critical density $n_c$, one of the
essential ingredients is the relevant degrees of freedom that
enter in the vicinity of the critical point. Depending upon what
enters there, certain aspects of the phase transition scenario can
be drastically different. These different scenarios will
eventually be sorted out either by experiments or by QCD
simulations on lattice or by both.

The standard way of addressing the problem at high temperature and
low density currently accepted by the majority of the community as
the ``standard picture" is to assume that near the critical point,
the only relevant low-lying degrees of freedom are the
(psudo-)Goldstone pions and a scalar meson that in the $SU(2)$
flavor case, makes up the fourth component of the chiral
four-vector of $SU(2)_L\times SU(2)_R$. For the two-flavor case,
one then maps QCD to an $O(4)$ universality class etc. Here
lattice measurements will eventually map out the phase structure.
On the contrary, at zero temperature and high density, the
situation is totally unclear. In fact as density increases, the
possibility is that one may not be able to talk about
quasipartcles of any statistics at all: The concept of a hadron
may even break down. Unfortunately lattice cannot help here, at
least for now, because of the notorious sign problem.

The situation is markedly different if the vector manifestation
\`a la Harada and Yamawaki~\cite{HY:VM,HY:PR} scenario is viable.
In this picture, certain hadrons other than pions can play a
crucial role in both high temperature and high density with a
drastically different phase structure. In particular, light-quark
vector mesons, i.e., the $\rho$ mesons,  can become the relevant
degrees of freedom near the phase transition, becoming ``sharper"
quasiparticles even near the critical density, thereby increasing
the number of degrees of freedom that enter from the hadronic
sector, with -- in contrast to the standard picture --
narrow-width excitations near the critical point. This can then
lead to a form of phase change that is a lot smoother than that of
the standard scenario.

This talk is a complement to the preceding talk by Masayasu
Harada.

\section{Hidden Local Symmetry and the Vector Manifestation}
To approach the chiral symmetry restored phase ``bottom up" from
the broken phase, we need an effective field theory (EFT) that
represents as closely as possible the fundamental theory of strong
interactions, QCD. In fact, according to Weinberg's unproven
theorem~\cite{wein-theorem},  QCD at low energy/momentum can be
encapsulated in an effective field theory with a suitable set of
colorless fields subject to the symmetries and invariance required
by QCD. So the question is how to construct an EFT that captures
as fully as possible the essence of QCD in describing the relevant
physics at the phase transition. In order to do this, we first
need to identify the scale at which we want to define our EFT and
the relevant degrees of freedom and symmetries that we want to
implement. I will consider two possibilities. One is the standard
scenario based on linear sigma model that assumes that the only
low-excitation degrees of freedom relevant to chiral restoration,
apart from the nucleons, are the pions and possibly a scalar
(denoted $\sigma$ in linear sigma model) with all other degrees of
freedom integrated out. The other is the vector manifestation
scenario based on hidden local symmetry (HLS) in which light-quark
vector mesons figure crucially.

In this talk, I will focus principally on what new physics can be
learned in the second scenario which seems to be currently
unappreciated by the community in the field. We consider for
definiteness the three-flavor case, i.e., $SU(3)_L\times SU(3)_R$.
The two-flavor case is a bit more subtle and has not yet been
fully worked out. I will leave out that issue. In going to nuclear
matter and beyond in this scenario, we must keep vector-meson
degrees of freedom $explicit$. This is because of the vector
manifestation (VM)~\cite{HY:VM,HY:PR} for which vector degrees of
freedom are indispensable. To understand the VM, we consider the
HLS Lagrangian~\cite{bandoetal} in which the pion and vector
mesons are the effective degrees of freedom. It should be stressed
that HLS is essential for the VM since local gauge symmetry is
required for doing a consistent chiral perturbation calculation in
the presence of vector mesons. Other theories where local gauge
symmetry is absent are moot on this issue. See \cite{HY:PR} for a
clear discussion on this point. For the moment, we ignore fermions
and heavier excitations as e.g., $a_1$, glueballs etc. To make the
discussion transparent, we consider three massless flavors, that
is, in the chiral limit~\footnote{Masses and symmetry breaking can
be introduced with attendant complications. Up to date, the
effects of quark masses have not been investigated in detail in
this formalism. It is possible that the detail structure of the
phase diagram can be substantially different from the chiral limit
picture we are addressing here.}. The relevant fields are the
(L,R)-handed chiral fields $\xi_{L,R}=e^{i\sigma/F_\sigma} e^{\mp
i\pi/F_\pi}$ where $\pi$ is the pseudoscalar Goldstone boson field
and $\sigma$ the Goldstone scalar field absorbed into the HLS
vector field $\rho_\mu$, coupled gauge invariantly with the gauge
coupling constant $g$. If one matches this theory to QCD at a
scale ${\Lambda_M}$ below the mass of the heavy mesons that are
integrated out but above the vector ($\rho$) meson mass, it comes
out -- when the quark condensate $\la \bar{q}q\ra$ vanishes as in
the case of chiral restoration in the chiral limit -- that
 \be
g(\bar{\Lambda})\rightarrow 0,\ \ a(\bar{\Lambda})\equiv
F_\sigma/F_\pi \rightarrow 1.
 \ee
Now the renormalization group analysis shows that $g=0$ and $a=1$
is the fixed point of the HLS theory and hence at the chiral
transition, one approaches what is called the ``vector
manifestation" fixed point. The important point to note here is
that {\it this fixed point is approached regardless of whether the
chiral restoration is driven by temperature $T$~\cite{HS:T} or
density $n$~\cite{HKR} or a large number of flavors~\cite{HY:NF}}.
At the VM, the vector meson mass must go to zero in proportion to
$g$, the transverse vectors decouple and the longitudinal
components of the vectors join in a degenerate multiplet with the
pions.
\section{Consequences on the Vector and Axial-Vector Susceptibilities}
As an illustration of what new features are encoded in the HLS/VM
scenario, we first consider approaching the critical point in heat
bath. We will come to the density problem later. Specifically
consider the vector and axial vector susceptibilities defined in
terms of Euclidean QCD current correlators as
 \be
\delta^{ab}\chi_V&=& \int^{1/T}_0 d\tau\int d^3\vec{x}\la V_0^a
(\tau, \vec{x}) V_0^b (0,\vec{0})\ra_\beta,\\
\delta^{ab}\chi_A&=& \int^{1/T}_0 d\tau\int d^3\vec{x}\la A_0^a
(\tau, \vec{x}) A_0^b (0,\vec{0})\ra_\beta
 \ee
where $\la \ra_\beta$ denotes thermal average and
 \be
V_0^a\equiv \bar{\psi}\gamma^0\frac{\tau^a}{2}\psi, \ \
A_0^a\equiv \bar{\psi}\gamma^0\gamma^5\frac{\tau^a}{2}\psi
 \ee
with the quark field $\psi$ and the $\tau^a$ Pauli matrix the
generator of the flavor $SU(2)$.
\subsection{\it The standard (linear sigma model) scenario}
The standard picture with pions figuring as the only degrees of
freedom in heat bath has been worked out by Son and
Stephanov~\cite{son}. The reasoning and the result are both very
simple and elegant. They go as follows.

If the pions are the only relevant degrees of freedom near the
chiral transition, then the axial susceptibility (ASUS) for the
system is encoded in the chiral Lagrangian of the form
 \be
{L}_{eff}=\frac{{f_\pi^t}^2}{4}\left({\Tr}\nabla_0 U\nabla_0
U^\dagger - v_\pi^2{\Tr}\del_i U\del_i U^\dagger\right) -\frac 12
\la\bar{\psi}\psi\ra {\rm Re}{\Tr} M^\dagger
U\label{LA}+\cdots\label{Leff}
 \ee
where $v_\pi$ is the pion velocity, $M$ is the mass matrix
introduced as an external field, $U$ is the chiral field and the
covariant derivative $\nabla_0 U$ is given by $\nabla_0 U=\del_0 U
-\frac i2 \mu_A (\tau_3 U +U\tau_3)$ with $\mu_A$ the axial
isospin chemical potential. The ellipsis stands for higher order
terms in spatial derivatives and covariant derivatives. Now given
(\ref{Leff}) as the $full$ effective Lagrangian which would be
valid if it could be given in a local form as is, then the ASUS
would take the simple form
 \be
\chi_A=-\frac{\del^2}{\del\mu_A^2} L_{eff}|_{\mu_A=0}={f_\pi^t}^2.
\label{chia}
 \ee
Within the scheme, this is the $entire$ story: There are no other
terms that contribute. That the ASUS is given solely by the square
of the temporal component of the pion decay constant follows from
the fact that the Goldstone bosons are the only relevant degrees
of freedom in the system, with those degrees of freedom integrated
out being totally unimportant. The effective theory of course
cannot tell us what $f_\pi^t$ is. However one can get it from
lattice QCD. To do this, we exploit that at the chiral
restoration, the vector correlator and the axial correlator must
be equal to each other, which means that
 \be
\chi_A|_{T=T_c}=\chi_V|_{T=T_c}.
 \ee
Now from the lattice data of Gottlieb et al~\cite{gottlieb}, we
learn that
 \be
\chi_V|_{T=T_c}\neq 0
 \ee
and hence from (\ref{chia}) that
 \be
f_\pi^t\neq 0.
 \ee
Next, we know that the space component of the pion decay constant
$f_\pi^s$ must go to zero at the chiral restoration. This is
because it should be related directly to the quark condensate
$\la\bar{q}q\ra$, i.e., the order parameter of the chiral symmetry
of QCD. Thus one is led to the conclusion that at $T=T_c$ the
velocity of the pion must be zero,
 \be
v_\pi\propto f_\pi^s/f_\pi^t\rightarrow 0 \ \ {\rm as} \ \
T\rightarrow T_c.
 \ee

That the pion velocity is zero at the critical point is analogous
to the sound velocity in condensed matter physics which is known
to go to zero on the critical surface. But the trouble is that
there is a caveat here which throws doubt on the simple result.
One might naively think that the vector susceptibility (VSUS)
could also be described by the same local effective Lagrangian but
with the covariant derivative now defined with the vector isospin
chemical potential $\mu_V$ as $\nabla_0 U=\del_0 U-\frac 12 \mu_V
(\tau_3 U-U\tau_3)$.  If the local form of the $effective$
Lagrangian (\ref{Leff}) is valid as well for the VSUS, one can do
the same calculation as for $\chi_A$, i.e.,
 \be
\chi_V=-\frac{\del^2}{\del\mu_V^2} L_{eff}|_{\mu_V=0}.
\label{chiv}
 \ee
Now a simple calculation shows that $\chi_V=0$ for {\it all T}.
This is at variance with the lattice result. It is also
unacceptable on general grounds. Thus either the local effective
Lagrangian is grossly inadequate for the VSUS or else the
assumption that the pions are the only relevant degrees of freedom
is incorrect. Indeed, Son and Stephanov suggest that diffusive
modes in hydrodynamic language that are not describable by a local
Lagrangian can be responsible for the non-vanishing VSUS.
\subsection{\it The HLS/VM scenario}
The situation is dramatically different in the VM
scenario~\cite{HY:VM,HY:PR}. The hidden gauge fields enter
importantly at the phase transition~\cite{HKRS}. The reason for
this is that their masses tend to zero near the VM fixed point and
hence they must enter on the same footing as the Goldstone pions.

In the HLS/VM scheme, the parameters of the effective Lagrangian
are defined at the matching scale $\Lambda_M$ in terms of the QCD
parameters that encode the vacuum change in heat bath and/or dense
medium. In computing physical observables like the current
correlators, one takes into account both quantum loop effects that
represent how the parameters run as the scale is changed from the
matching scale to the physical (on-shell) scale $and$ thermal
and/or dense loop effects induced in the renormalization-group
flow. Now the former implies an $intrinsic$ temperature and/or
density dependence in the parameters~\footnote{This dependence is
missing in most of the effective field theory calculations that
are based on effective Lagrangians determined in the matter-free
and zero temperature vacuum. Most of the treatments found in the
literature nowadays belong to this category.} -- called
``parametric dependence." The one-loop calculations in \cite{HKRS}
show that both effects are governed by the VM fixed point at
$T\rightarrow T_c$,
 \be
g\rightarrow 0, \ \ a\rightarrow 1.
 \ee
The results~\cite{HKRS,HKRS2} that follow from this consideration
are
 \be
f_\pi^t|_{T=T_c}=f_\pi^s|_{T=T_c}=0, \ \ v_\pi|_{T=T_c} \lsim
1\label{main1}
  \ee
and~\footnote{In the original version of Harada, Kim, Rho and
Sasaki~\cite{HKRS,MRyukawa}, quasiquarks were introduced near the
critical point. I think there is a bit of over-counting here. In
fact it seems more appropriate to simply drop the quasiparticle
contributions all together. One can justify this by arguing that
one should be introducing color-singlet fermions, namely, baryons
rather than the colored quasiparticles which are not physical. Now
the baryons can be considered to become light near the phase
transition in the spirit of BR scaling~\cite{BR91} but stay
heavier than the hidden gauge bosons, so one can imagine
integrating them out along with other heavier hadrons such as the
$a_1$ mesons, scalars etc., thus preserving the same degrees of
freedom near the critical point as in zero-temperature space. If
baryons were included in the description, then it would be
necessary to assure self-consistency between
baryon-particle-baryon-hole configurations (called ``sobars") and
the elementary mesons that can be mixed in medium.}
 \be
\chi_A|_{T=T_c}=\chi_V|_{T=T_c}\approx  \frac{N_f^2}{6} T_c^2  \
\label{main2}
 \ee
which is more or less what was found in the lattice QCD
calculation~\cite{gottlieb,BR02}. One can understand these results
as follows. As $T\rightarrow T_c$, both $f_\pi ^{t,s}$ approach
$\bar{f}_\pi\propto \la\bar{\psi}\psi\ra$ which approaches
zero~\cite{HS:T}. At one loop, they approach the latter in such a
way that the ratio goes near (but not quite) 1, thereby making the
pion velocity approach near the velocity of light~\footnote{Due to
Lorentz-symmetry breaking by medium, there is a small
deviation~\cite{HKRS2}, say, $\sim 15 \%$, from 1 in the
parametric pion velocity in the ``bare" Lagrangian at the matching
scale. The quantum correction to this is protected by the VM, so
the deviation remains unchanged in the flow of RGE.}. Both
$\chi_{V,A}$ get contributions from the flavor gauge vector mesons
whose masses approach zero and chiral symmetry forces them to
become equal to each other. {\it Since both the space and time
components of the pion decay constant are vanishing, they do not
figure in the formulas for the susceptibilities (\ref{main2}) in
sharp contrast to the result of Son and Stephanov~\cite{son}}.

If realized, the VM scenario will present an interesting phase
structure. It will give a phase diagram drastically different from
the standard sigma model one. For instance, it would imply that
there are a lot more degrees of freedom than in the standard
picture just below the critical temperature. We do not know how
fast the masses actually drop as one approaches, bottom-up, the
critical point but if the presently available lattice results are
taken at their face value, then they do not seem to drop
appreciably up to near $T_c$. But it is still possible that they
drop to zero in a narrow window near the critical point in a way
consistent with the VM and account for the rapid increase of
energy density observed in the lattice calculations. In any event,
that would provide a natural explanation of a smooth transition
with a possible coexistence of excitations of various quantum
numbers below and above $T_c$ as seems to be indicated by the MEM
analysis of Asakawa, Hatsuda and Nakahara~\cite{hatsuda}.

I must mention that there is a caveat here. The results
(\ref{main1}) and (\ref{main2}) are one-loop results and one may
wonder whether two-loop or higher orders -- which are presently
too laborious to compute -- would not change the qualitative
features. The space component of the pion decay constant is
undoubtedly connected to the chiral order parameter, so remains
zero to all orders but there is no general argument to suggest
that the time part cannot receive non-vanishing contributions at
higher orders. If it did, then we would fall back to the
Son-Stephanov result of a vanishing pion velocity.
\section{Multiplet structure}
The multiplet structure of hadrons implied by the VM at the phase
transition at $T=T_c$ or $n=n_c$ or $N_f=N_f^c$ is basically
different from that of the standard one based on linear sigma
model. Continuing with three flavors in the chiral limit, at the
phase transition where the VM is realized, the longitudinal
components of the vector mesons $\rho_\parallel$ join the
$(1,8)\oplus (8,1)$ multiplet of the Goldstone pions and the
transverse vectors, massless, decouple from the system as the
gauge coupling vanishes. This contrasts with the linear sigma
model picture where a scalar joins the Goldstone pions in
$(3,3^*)\otimes (3^*,3)$ multiplets. If one were to explicitly
incorporate the $a_1$ vector mesons and the scalar meson in the
scheme (so far integrated out), then the $a_1$ would be in the
same multiplet $(3,3^*)\otimes (3^*,3)$ with the scalar in the VM
while in the standard picture, the $a_1$ would be in the multiplet
$(1,8)\oplus (8,1)$ with the vectors $\rho$.

As stressed by Harada and Yamawaki, physically there is no sense
to be on the VM point. It makes sense only to $approach$ it from
below. One can however ask what happens precisely at the VM. Since
the VM is in the Wigner mode, one may wonder whether the decoupled
vector mesons should not have chiral partners. If they should,
then the HLS/VM would be in difficulty since the theory does not
contain such chiral partners in the same multplet. The only way
that I see to avoid this obstruction is that the transverse
vectors become ``singlet" under chiral transformation. This can be
made possible if the chiral transition is considered as the flavor
vector mesons getting de-Higgsed to the color gauge bosons, i.e.,
gluons, as proposed by Wetterich~\cite{wetterich}. Specifically
one can think of the flavor vector mesons as the vectors excited
in the color-flavor locking (CFL) transition
 \be
SU(3)_L\times SU(3)_R\times SU(3)_c\rightarrow SU(3)_{L+R+c}
 \ee
in analogy to the CFL in color superconductivity in QCD at
asymptotic density. In this case, the flavor vectors un-lock the
color and flavor and turn in a sort of relay~\cite{MR:Taiwan} into
the color gauge vector mesons. This phenomenon can be summarized
by writing
 \be
\xi_{L(R)i}^\alpha=[\xi_{L(R)}v]_i^a, \ \ U=\xi_L^\dagger\xi_R
 \ee
in terms of a color-singlet $\xi$ field and a $v\in SU(3)_C$, both
of which are unitary. One can then
relate~\cite{wetterich,MR:Taiwan} the vector meson fields
$\rho_\mu$ and the baryon fields $B$, to the gluon fields $A_\mu$
and the quark fields $\psi$ as
 \be
B&=&Z_\psi^{1/2}\xi^\dagger \psi v^\dagger,\nonumber\\
\rho_\mu&=&v (A_\mu  +\frac{i}{g} \del_\mu)
v^\dagger\label{dressed}
 \ee
where $Z_\psi$ is the quark wave function renormalization
constant. This CF unlocking scenario appears to be highly
appealing and fascinating. Up to date, however, this idea has not
been fully worked out -- the mechanism for CFL and CF-unlocking is
not known -- and although quite plausible, it is not proven yet
that it is not inconsistent with the known structure of QCD. It
remains to be investigated.
\section{Evidences for the VM?}
The VM prediction is clean and unambiguous for $SU(3)_L\times
SU(3)_R$ chiral symmetry in the chiral limit at the chiral
restoration point. At present, there are no lattice measurements
that validate or invalidate this picture. Are there experimental
indications that Nature exploits this scheme?

So far, there are no indications from relativistic heavy-ion
processes as to whether the VM is realized in the vicinity of the
critical temperature or density. So to answer this question, one
would have to work out what happens at temperatures and/or
densities away from the critical point. One would also have to
consider two-flavor cases and quark mass terms to make contact
with Nature. To do all these is a difficult task and no
theoretical work has been done up to date on this matter.  What is
available up to date are some indications in nuclear systems at
low temperature. Nuclei involve many nucleons in the vicinity of
nuclear matter density and the density regime involved here is
rather far from the density relevant to the VM. Thus we are
compelled to invoke a certain number of extrapolations toward the
VM point. Near nuclear matter density, however, we have a
many-body fixed point known as ``Fermi-liquid fixed
point"~\cite{FR,shankar,MR:Taiwan} which involves quantum critical
phenomenon and this makes the connection to the VM tenuous even
when one is at a much higher density.

To make progress in this circumstance, we have to make a rather
drastic simplification of the phase structure. Here we will adopt
what is called ``double-decimation approximation"~\cite{BR-DD}
which consists of (1) extrapolating downwards from the VM to the
Fermi-liquid fixed point and (2) extrapolating upwards from the
zero-density regime where low-energy theorems apply to the
Fermi-liquid fixed point at nuclear matter density. The spirit
here is close to BR scaling~\cite{BR91} proposed in 91.

Close to the VM, the vector meson mass must go to zero in
proportion to $g$.  Specifically\footnote{Here and below, I denote
the quark field by $q$ instead of $\psi$ used before.}
 \be
m_\rho^*/m_\rho \approx g^*/g \approx
\la\bar{q}q\ra^*/\la\bar{q}q\ra \rightarrow 0
 \ee
as the transition point $n=n_c$ is reached. One can understand
this as follows. Near the critical point the ``intrinsic term"
$\sim g^* F_\pi^*$ in the vector mass formula drops to zero faster
than the dense loop term that goes as $\sim g^* H(n)$ where $H$ is
a slowly (i.e., logarithmically) varying function of density. So
the dense loop term controls the scaling. Now it seems to be a
reasonable thing to assume that $near$ the VM fixed point, we have
the scaling
 \be m_\rho^*/m_\rho \approx g^*/g\approx
\la\bar{q}q\ra^*/\la\bar{q}q\ra.
 \ee
Our conjecture~\cite{BR-DD} is that this holds down to near
nuclear matter density.

Let us now turn to the low-density regime, that is, a density
below nuclear matter density. At near zero density, one can apply
chiral perturbation theory with a zero-density HLS Lagrangian
matched to QCD at a scale $\Lambda_M\sim \Lambda_\chi$. We expect
to have~\cite{BR02}
 \be
m_\rho^*/m_\rho \approx f_\pi^*/f_\pi \approx
\sqrt{\la\bar{q}q\ra^*/\la\bar{q}q\ra}.\label{dd1}
 \ee
This result follows from an in-medium GMOR relation for the pion
if one assumes that at low density the pion mass does not scale
(as indicated experimentally~\cite{yamazaki}), that the vector
meson mass is dominantly given by the ``intrinsic term" $\sqrt{a}
F_\pi g$ with small loop corrections that can be ignored and that
the gauge coupling constant does not get modified at low density
(as indicated by chiral models and also empirically). The
double-decimation approximation is to simply assume that this
relation holds from zero density up to nuclear matter
density~\cite{BR-DD}. Note that we are essentially summarizing the
phase structure up to chiral restoration by two fixed points,
namely, the Fermi-liquid fixed point and the vector-manifestation
fixed point. Here we are ignoring the possibility that there can
be other phase changes, such as kaon condensation (or hyperon
matter), color superconductivity etc. which can destroy the Fermi
liquid structure before chiral symmetry is restored.

The one important feature that distinguishes the HLS/VM theory
from other EFTs is the $parametric$ dependence on the background
of the ``vacuum" -- density and/or temperature -- which
intricately controls the fixed point structure of the VM. At low
density, this dependence is relatively weak, so hard to pinpoint.
But in precision experiments, it should be visible. One such case
is the recent experiment of deeply bound pionic atoms. For this,
we can consider a chiral Lagrangian in which only the nucleon and
pion fields are kept explicit with the vectors and other heavy
hadron degrees of freedom integrated out from the HLS/VM
Lagrangian. The relevant parameters of the Lagrangian are the
``bare" nucleon mass, the ``bare" pion mass, the ``bare" pion
decay constant, the ``bare" axial-vector coupling and so on which
depend non-trivially on the scale $\Lambda_M$ and density $n$.
This Lagrangian takes the same form as the familiar one apart from
the $intrinsic$ dependence of the parameters on $n$. (In the usual
approach, the scale $\Lambda_M$ is fixed at the chiral scale and
the dependence on $n$ is absent). As shown by Harada and
Yamawaki~\cite{HY:matching,HY:PR}, the local gauge symmetry of HLS
Lagrangian enables one to do a systematic chiral perturbation
theory even when massive vectors are present. Since the vectors
are integrated out, the power counting will be the same as in the
conventional approach. Now if the density involved in the system
is low enough, say, no greater than nuclear matter density, then
one could work to leading order in chiral expansion. Suppose that
one does this to the (generalized) tree order. To this order, the
parameters of the Lagrangian can be identified with physical
quantities. For instance, the bare pion decay constant $F_\pi$ can
be identified with the physical constant $f_\pi$, the parametric
pion mass with the physical pion mass $m_\pi$ etc. Now in the
framework at hand, the only dependence in the constants on density
will then be the $intrinsic$ one determined by the matching to QCD
immersed in the background of density $n$.

If we apply the above argument to the recent measurement by Suzuki
et al~\cite{yamazaki} of deeply bound pionic atom systems, we will
find that the measurement supplies information on the ratio
$f_\pi^*/f_\pi$ at a density $n\lsim n_0$. There is a simple
prediction for this quantity~\cite{FR,BR:PR01}. We have from
(\ref{dd1})
 \be
\Phi (n)\equiv f_\pi^*/f_\pi \approx
\sqrt{\la\bar{q}q\ra^*/\la\bar{q}q\ra} .\label{dd2}
 \ee
Instead of calculating the quark condensate in medium which is a
theoretical construct, we can extract the in-medium pion decay
constant by extracting $\Phi$ from experiments. Indeed, the
scaling $\Phi$ has been obtained from nuclear gyromagnetic ratio
in \cite{FR,BR:PR01}. At nuclear matter density, it comes out to
be
 \be
\Phi (n_0)\approx 0.78
 \ee with an uncertainty of $\sim 10\%$. Thus it is predicted that
 \be
(f_\pi^*(n_0)/f_\pi)_{th}^2 \approx 0.61.\label{fpith}
 \ee
This should be compared with the value extracted from the pionic
atom data of \cite{yamazaki},
 \be
(f_\pi^*(n_0)/f_\pi)_{exp}^2=0.65\pm 0.05.
 \ee

It is perhaps important to stress that this ``agreement" cannot be
taken as an evidence for ``partial chiral restoration" as one
often sees stated in the literature. Apart from ambiguity in
interpreting the experimental results, in particular, in how the
order parameter of chiral restoration is extracted from the
experimental data, there is also a theoretical ambiguity. For
instance, if one were to go to higher orders in chiral expansion,
the $parametric$ pion decay constant cannot be directly identified
with the physical pion decay constant since the latter should
contain two important corrections, i.e., quantum corrections
governed by the renormalization group equation as the scale is
lowered from $\Lambda_M$ to the physical scale and dense loop
corrections generated by the flow. At the chiral restoration, it
is this latter that signals the phase transition: The parametric
pion decay constant with the scale fixed at the matching scale
does not go to zero even at the chiral restoration
point~\cite{HY:PR}. Thus when one does a higher-order chiral
perturbation calculation of the same quantity, one has to be
careful which quantity one is dealing with.

What one can say with some confidence is that (\ref{fpith}) goes
in the right direction in the context of BR scaling~\cite{BR91}.

A variety of other evidences that lend, albeit indirect, support
to the scaling~\cite{BR91} and in consequence to the notion of the
VM are discussed in \cite{BR:PR01,BR-DD,LPRV}. If the VM were
verified by going near the chiral transition point, it would
constitute a nice illustration of how the mass of the hadrons
making up the bulk of ordinary matter around us is made to
``disappear," a deep issue in physics~\cite{wilczek}.
\subsection*{Acknowledgments}

It gives me an immense pleasure to give a talk on the hidden local
symmetry approach to hadronic physics pioneered by the chair of
this meeting, Koichi Yamawaki, and his colleagues. I have always
been strongly influenced by the power and elegance of this
approach. I am grateful for numerous discussions with Gerry Brown,
Masayasu Harada, Youngman Kim and Koichi Yamawaki.

\end{document}